# Intelligent Load Balancing Systems using Reinforcement Learning System

Raju Singh
*Arizona State University, US*
singh.raju@msn.com

*Abstract*—Load Balancing is a critical technological advancement for scaled cloud infrastructure. It allows the systems to distribute the load to the backend servers with pre-defined algorithms such as round robin, weighted round-robin, least connection, weighted least connection, resource based (adaptive), weighted response time, source IP hash, url hash etc. With this approach, software developers, infrastructure administrators and systems admins have been able to address the most of the Internet traffic related issues for load balancing on contemporary software stacks such as – monolith, classical three tier architecture, microservices architecture etc. However, we can assume that traffic balancing techniques may soon be inefficient for enabling optimized distribution time, as existing algorithms failed to meet the Internet traffic requirements going forward, and hence result in poor user experience. To address the traffic distribution need (in terms of response time, distribution time, uptime etc.), we must think to address this problem in advance. The key problem faced by load balancers are many, a few of it are – traffic management, congestion handling, scheduling, know when to load balance and when not to.

## I. INTRODUCTION

In recent time, most of the on-prem application hosting has moved to cloud. Cloud is a diversified field, a proven strong contender in terms of catering application across the globe over the web. However, this has led to several challenges at each step of hosting. Software and hardware scientists are working to remediate challenges posed by exponential growth of Internet by carefully and scientifically examining the problem and devising new approaches. One such problem statement is related to load balancing. Load balancing has been used since the time of Internet evolution. To optimally serve, millions and millions of customers across different geographies, service provider companies have started to think about efficient, distributed, yet robust and unified system which can distributed millions of web requests to thousands of backend servers. One of the most important aspect to be highlighted here is – understanding the real computational science behind load balancing and why it is needed. Load balancing roots back to the problem of scheduling jobs on identical nodes, however, the it definition has expanded its founding aspect, as its mostly commonly described as:

Load balancing is a technique to divide the workload onto multiple resources in order to avoid overload on any of the resources . Maximizing throughput, minimizing response time, latency and optimizing traffic are some of the load balancing goals . Cloud networks have a global view of the network. So optimizing the web traffic management using load balancer is the key to business success.

## II. BACKGROUND

### A. Aim of the Project

The principal aim of the project is - how to architect, design, develop and implement web traffic distribution algorithm which can meet the near real time user experience and adjust the load on the backend servers by analysing inputs intelligently from several sources, such that no servers in the target group has skewed request assignment (one of the several criteria). The problem becomes more complex to improvise as we maintain different servers with different capacity in terms of processing power. Load balancing also suffers from fundamental issues like scheduling, congestion, and mean time to distribute the load while not compromising the user experience in terms of response time, latency etc. There are several salient features to the algorithm. It makes use of in memory queues, machine learning model, health check systems etc. This way, the proposal looks at the holistic approach and solves few fundamental shortcomings of contemporary load balancing techniques

### B. Motivation

Since the time of cloud evolution, computer scientist has been consistently facing the heat to improve collective performance of the cloud resources. Underutilization of cloud resources causes financial loss to the service provider as cloud is mostly a pay-as-use model. A solution to - the increased complexity of server farms, and the need to improve the server utilization is to implement load balancing technique. It consists of a front-end machine which intelligently redirects the traffic to several backend servers. The motivation behind analysing the load balancing problem comes from the recent developments in the area of traffic analysis and routing techniques. Consistent efforts being made to facilitate optimized load balancing solution. Example:- consider a use case with three elements - users, load balancer and servers. The user consumes the resources over web, using an end-point. When users request for resources using the URI, it hits the load balancer, which then performs some pre-requisites and then it distributes the load to the backend servers using as per LB algorithm. New generation of load balancer supports auto-scaling, which use the load statistics and scale-up/scale-down features. It also support for burst type load distribution.

Efforts has been made to overcome these shortcomings such as efficient response time, traffic scheduling, few of efforts are impressive scientific work, however, few are more of work around. Web cache is a technique to serve the client with requested resource, but it must not be confused with load balancing category, but rather it is cache technique. Web caching also suffers from the problem of cache invalidation, when it comes to an ever changing data.

Let us first understand what is a load balancer and how it addresses the need for modern business application users.

Load Balancing - A pool (called server farm) of identical servers is connected with the Internet by a load balancer that acts as a front-end machine which intelligently directs the clients requests (their TCP connection) to the backend servers according to the servers' capacities and status. This allows each server to operate more efficiently. The choice of the backend server falls into two basic categories: **content unaware** and **content aware**.

**Content Unaware**: There are no dedicated servers for some specific protocols and thus all of them can process any kind of request. Every time that the load balancer receives a request of connection (SYN), it redirects the TCP connection to a backend server which will finish the three-way handshake. The redirection is generally based on Layer 2, 3 or 4 information such as IP addresses or ports numbers.

**Content Aware**: The load balancer is an end-point of the TCP connection, and can understand the user's request through a parsing of layer-7 information (i.e. URL). The redirection happens after the three-way handshake between the load balancer and the user. Now all servers can have different contents or different levels of security or power.

The later solution is more flexible but less fast than the previous one considering that the load balancer must be able to parse layer-7 information spread over several packets. Another advantage of later solution is that is able to manage sticky connections. Some applications require that some user connections are always redirected to the same real server. Examples are filling of forms, shopping carts, bank transactions, payments, etc. These particular connections are typically handled in three ways:

- The load balancer stores the user and source IP information into the table. This solution does not work with the Network Address Translation (NAT) because for each new connection of the same user, there is a different, and dynamically assigned, source IP address
- The server or the load balancer analyzes sticky connection and sends a cookie back to the user that will be used next time to know where to redirect the connection. It does not work with encrypted connections.
- The server or the load balancer uses HTTP to redirect to the correct real server.

However, this system suffers from drawback of:

- Uneven request assignment to target servers
- Traffic distribution rules are mostly static and pre-defines with little to no scope of handling business use cases.
- Limited load balancing algorithms, inefficient for a large-scale distributed infrastructure
- Scheduling (knowing when and how to load balance optimally) and congestion problem.

Several approaches has been made to address the load balancing issues, however each approach has its own pros and cons. A new approach, enabled with machine learning algorithm can enhance the traffic distribution systems by:

- introducing a queuing mechanism at the load balancing tier. This would allow the load balancer to queue the incoming traffic, store it in different priority queues. The queues act as placeholder, which handles extreme traffic scenarios in several ways like - not returning 'Server Busy', 'Request time out', latency issues or any other server errors to the requester, etc.

- supervising target server system externally using a machine learning model (RL). This system gathers statistics of the server such as request handled successful, return codes, system statistics etc. The information is passed to a system, which analyzes the target server health and fed the info to the ML system.

The queuing system queues up the incoming request in parallel in-memory (multiple) queues. The target groups request the queues (using client-side of application, running on the target group server). The target system is supervised by a machine learning model (RL), which credits cpu/memory cycle the target groups which processes request progressively.

### III. RESEARCH APPROACH

The proposed research methodologies is a culmination of several technologies, including reinforcement learning, queue management, API functions etc. It walks us through different phases of load balancing, describe a potential problem and try to obtain a solution which has practical implications. Let us assume a system, which has three components (at a basic level). The **load-balancer**, the **target groups servers,** and the **reinforcement learning model**.

**Load-balancing tier**: User requests (Web requests) hits the load balancer, which is placed in between the end user and the target servers. The load balancer checks the request types, and identifies the request resources by analyzing the request. Each request type has attributes and can **GET**, **POST**, **PUT**, **Email**, **Chat**, **upload**, **download**, **SYNC** etc. The load balancer tier also consists of a host of multiple in-memory queues. The load balancer channelizes the request to the queues based on its attributes, and classification rules. These classification rules are static and defined at once while designing the load balancer.

**Target Group tier**: The backend server or server farms, which runs specialized application in order process the incoming requests, for example, an email handles, a file uploader, a sync daemon, api handler etc. This tier has several nodes, on each of its nodes runs a client side of the queue application. It is important to note that queue implementation on the load balancing. This client queue software which is running on each server in backend server farm, reads the respective LB queues for requests and pulls the queued request on the client nodes.

**Reinforcement Learning** Model is a separate system, outside the load balancing tier and target group server tier. Reinforcement Learning supervises the agents (queue client software) running on the target group server. The intelligent actions can be thought of a way to pull (action) the queued requests from the load balancer tier within a stipulated time. The reward is in the form of cpu/memory cycles, if the request pull on client server meets the qualifying criteria.

**Reinforcement learning** [def]
Reinforcement learning (RL) is a general framework for building autonomous agents (physical or virtual), which are systems that make decisions without human supervision in order to perform a given task. For example, In all those examples, an agent faces a sequential decision-making problem, which can be represented as an interaction loop between an agent and an environment. After observing its current situation, the agent selects an action to perform. As a result, the environment changes its state and provides a numeric reward feedback about the chosen action. In RL, the agent needs to learn how to choose good actions based on its observations and the reward feedback, without necessarily knowing the dynamics of the environment.
.
In such problems, an agent faces a sequential decision-making problem where, at every time step, it observes its state, performs an action, receives a reward and moves to a new state. An RL agent learns by trial and error a good policy (or controller) based on observations and numeric reward feedback on the previously performed action.

**RL system interaction with Load Balancing System**:
RL System tier interacts with the pool of resources (CPU/memory) over an API call. It does not reward the servers from the farm, when the action to pull the request from the load balancing queues exceed the stipulated time under observation. This encourages the system to keep the request pull time within limited, and the system keep improving with additional rewards.

Pseudo Steps

- User sends the request to access the resources using the web URI.
- Requests hits the load balancer, SSL offload is performed (in case of secured traffic), requests gets queued up based on the request type. The decision to queue the request is done by the algorithm at the load balancer tier.
- Target group runs intelligent agents (clients) – queue agent (which keeps checking the queue status on the load balancer – queue depths) and the RL agent (which reports to RL system, as how many requests are being processed by the server, how much time it took to process each requests etc.).
- The queue client agent's actions request pulls from the queues at the load balancing tier.
- RL system monitors the agents and reward/un-rewards it's action by granting/dis-granting cpu/memory cycles.
- The cpu/memory is made available to the server as the server process optimally requests successively. However, this is not an unwarranted addition of the resources to the server. The RL system keeps a track of the total rewards to a server and it stops once server reaches a threshold.
- On the other hand, the server which shows sign of degradation, are un-rewarded with cpu/memory credit and finally reaches a state where the system discards the server from the web farm.

## IV. ACKNOWGEMENT

Few things which are not part of this proposal:
*Scaling Load balancer*: As the number of web requests hits the load balancers, the load balancer should adjust or recalibrate itself in terms of resources assigned to it.

*Congenstion management*: Congestion occurs due to many reasons, deficiency of resources and irregularity in resource allocation. The stereophonic methods for congestion control such as rate control, window mechanism, queue control and others can be executed to solve the congestion problem, but for the congestion due to irregular allocation, the fundamental solution is to use network resources more effectively by adjusting the traffic routing depending on choice may be probabilistic, when congestion occurs. Congestion on network creates delay in transmission of data which also leads to loss of the data packets, wastage of time and decrease in lifetime of the network channel. Main goal of our proposed algorithm is to find out the areas with congestion between source and destination nodes which will be helpful in avoidance of the congestion on the network in intermediate links which also reduces the packet loss on the communication channel.

REFERENCES

[1] Berenbrink, P., Friedetzky, T., Kaaser, D. and Kling, P., 2019, May. Tight & simple load balancing. In 2019 IEEE International Parallel and Distributed Processing Symposium (IPDPS) (pp. 718-726). IEEE.

[2] Assadi, S., Bernstein, A., & Langley, Z. (2020). Improved bounds for distributed load balancing. arXiv preprint arXiv:2008.04148.

[3] Berndt, S., Deppert, M. A., Jansen, K., & Rohwedder, L. (2022). Load Balancing: The Long Road from Theory to Practice. In 2022 Proceedings of the Symposium on Algorithm Engineering and Experiments (ALENEX) (pp. 104-116). Society for Industrial and Applied Mathematics.

[4] Zhu, X., Zhang, Q., Cheng, T., Liu, L., Zhou, W., & He, J. (2021, September). DLB: Deep Learning Based Load Balancing. In 2021 IEEE 14th International Conference on Cloud Computing (CLOUD) (pp. 648-653). IEEE.

[5] Berenbrink, P., Friedetzky, T., Kaaser, D., & Kling, P. (2019, May). Tight & simple load balancing. In *2019 IEEE International Parallel and Distributed Processing Symposium (IPDPS)* (pp. 718-726). IEEE.

[6] Assadi, S., Bernstein, A., & Langley, Z. (2020). Improved bounds for distributed load balancing. *arXiv preprint arXiv:2008.04148*.

[7] Berndt, S., Deppert, M. A., Jansen, K., & Rohwedder, L. (2022). Load Balancing: The Long Road from Theory to Practice. In *2022 Proceedings of the Symposium on Algorithm Engineering and Experiments (ALENEX)* (pp. 104-116). Society for Industrial and Applied Mathematics.

[8] Boulmier, A., Abdennadher, N., & Chopard, B. (2021). Optimal Load Balancing and Assessment of Existing Load Balancing Criteria. *arXiv preprint arXiv:2104.01688.*